\documentclass[twocolumn, twocolappendix]{aastex631}

\usepackage{amsmath,amstext}
\usepackage{CJK}
\usepackage[T1]{fontenc}

\usepackage{newtxtext,newtxmath}
\usepackage[figure,figure*]{hypcap}
\usepackage{booktabs}
\usepackage{cleveref}
\usepackage{paralist}
\graphicspath{{fig/}}
\usepackage{ulem}
\usepackage[shortlabels]{enumitem}
\usepackage{xcolor,colortbl}  

\usepackage[colorlinks=true]{hyperref}

\newcommand{\mix}{\mathrm{mix}}

\newcommand{\gq}{\gamma_q}

\newcommand{\amf}{\alpha_\mathrm{mf}}

\shorttitle{MiMO-catalog}
\shortauthors{Li et al.}

\begin{document}
\begin{CJK*}{UTF8}{gbsn}

\title{The MiMO Catalog: Physical Parameters and Stellar Mass Functions of 1,232 Open Clusters from Gaia DR3}

\correspondingauthor{Lu Li}
\email{lilu@shao.ac.cn}

\author[0000-0002-0880-3380]{Lu Li (李璐)}
\affil{Shanghai Astronomical Observatory, Chinese Academy of Sciences, 80 Nandan Road, Shanghai 200030, China}

\author[0000-0001-8611-2465]{Zhengyi Shao (邵正义)}
\affil{Shanghai Astronomical Observatory, Chinese Academy of Sciences, 80 Nandan Road, Shanghai 200030, China}
\affil{Key Lab for Astrophysics, Shanghai 200234, China}

\author[0000-0001-7890-4964]{Zhaozhou Li (李昭洲)}
\affil{School of Astronomy and Space Science, Nanjing University, Nanjing, Jiangsu 210093, China}
\affil{Key Laboratory of Modern Astronomy and Astrophysics, Nanjing University, Ministry of Education, Nanjing 210093, China}
\affil{Centre for Astrophysics and Planetary Science, Racah Institute of Physics, The Hebrew University, Jerusalem, 91904, Israel}

\author[0000-0002-6506-1985]{Xiaoting Fu (符晓婷)}
\affil{Purple Mountain Observatory, Chinese Academy of Sciences, Nanjing 210023, China}

\begin{abstract}
We present a homogeneous catalog of 1,232 open clusters with precisely determined ages, metallicities, distances, extinctions, and stellar mass function (MF) slopes, derived from Gaia DR3 data. The parameters are inferred using the Mixture Model for Open clusters (MiMO), a novel Bayesian framework for modeling clusters in the color-magnitude diagram. By explicitly accounting for field-star contamination as a model component, MiMO removes the conventional need for stringent membership preselection, allowing for a more complete inclusion of member stars and thereby enhancing both precision and robustness. Our results broadly agree with existing catalogs but offer improved precision. For each cluster, we provide the best-fit age, metallicity, distance, extinction, and MF slope, along with their full likelihood chains and photometric membership probabilities for individual stars. We further identify an ``MF Prime'' subsample of 163 clusters with high-quality data, for which the MF estimates are considered most reliable.
The catalog and an open-source implementation of MiMO are made publicly available to the community.
\end{abstract}


\defcitealias{Li2022k}{LS22}
\defcitealias{Dias2021a}{D21}

\section{Introduction} \label{sec:intro}

Star clusters are not only fundamental laboratories for studying stellar formation and evolution, but also key components in galaxy formation and evolution. In particular, open clusters (OCs), which primarily reside in the Galactic disk, exhibit a wide range of ages and metallicities, making them excellent tracers of the disk's structure and evolution history \citep{Becker1970,Janes1982,Cantat-Gaudin2020a,Castro-Ginard2021a}. A comprehensive catalog of OC parameters can provide crucial insights into the processes of stellar formation and dynamical evolution.

Two pioneering and widely used catalogs of OC parameters before the Gaia era were presented by \citet{Kharchenko2005} and \citet{Dias2002,Dias2012}. The advent of the Gaia mission, with its unprecedented precision in photometric and astrometric measurements, has brought the renaissance of OC studies. Numerous new OC candidates have been discovered, nearly doubling their number \citep{Castro-Ginard2019a,Liu2019d,Sim2019a,hunt2023,perren2023}.

In parallel, several methods have been developed to infer OC parameters from color-magnitude diagrams (CMDs). Many of these are based on isochrone fitting, utilizing various optimization techniques, such as the cross-entropy method \citep{Dias2012,Dias2021a} or Bayesian frameworks like ASteCA \citep{Perren2015,Perren2022a} and BASE-9 \citep{vonhippelInvertingColorMagnitudeDiagrams2006,vonhippelBayesianAnalysisStellar2014,Bossini2019a}. More recently, deep learning approaches have also been employed to estimate cluster parameters \citep{Cantat-Gaudin2020b,hunt2023,cavalloParameterEstimationOpen2024}.

Most existing OC catalogs estimate parameters by comparing the distribution of member stars to theoretical isochrones in the CMD, effectively treating both as curves. The shape of the isochrone is determined by age, metallicity and extinction, while its position in the CMD is further shifted by distance.

However, as extensively discussed in \citet[hereafter \citetalias{Li2022k}]{Li2022k}, traditional CMD fitting approaches face major limitations. A common challenge is the inherent trade-off in membership selection. For instance, methods that rely on a pre-selected sample of high-probability members, such as in \citet{Kharchenko2013a}, often employ strict criteria to ensure high purity, but this can inadvertently remove key member stars. While some modern methods, like the isochrone fitting frameworks ASteCA \citep{Perren2015,Perren2022a} and certain deep learning models (e.g., \citet{Cantat-Gaudin2020a}), mitigate this by handling field star contamination probabilistically, many curve-based fitting methods do not exploit the full information from data, such as the stellar mass function, binary fraction, or binary mass ratio distribution.

To address these challenges, \citetalias{Li2022k} developed a novel Bayesian framework, the Mixture Model for Open Clusters (MiMO). MiMO models the CMD as a probabilistic mixture of single and binary cluster members and field stars. This approach removes the need for stringent membership selection in traditional methods, allowing for more inclusive sample selection and more precise and robust parameter estimation.

 As a rigorous Bayesian model, MiMO enables precise inference of key cluster properties, including isochrone parameters (age, distance, metallicity, and extinction). Moreover, MiMO also fits properties of the stellar mass distribution, such as the stellar mass function and binary population parameters (binary fraction and mass ratio distribution), which are inaccessible to conventional isochrone fitting methods. 

However, it is important to note that in the present work, due to known discrepancies between theoretical isochrone models and observed main sequences, binary star properties inferred without first correcting these discrepancies would be unreliable. Therefore, in this study, the binary fraction is treated as a nuisance parameter rather than a physically meaningful result, and the slope of the binary mass ratio distribution is fixed (as detailed in Section 2.3); thus, these binary parameters are not reported as primary results in the final catalog. \citetalias{Li2022k} has validated the accuracy and applicability of the method through extensive mock tests.

In this work, we apply MiMO to a large sample of known OCs, producing a homogeneous catalog for 1232 clusters. For each cluster, we provide not only the fundamental physical parameters but also the stellar mass function slope. This catalog is based on Gaia DR3 photometric data, without pre-selection of member stars. From this full catalog, we identify an ''MF Prime sample'' of 163 clusters. These clusters, selected for their superior data quality, form a robust dataset that can serve as a solid foundation for future detailed studies on the evolution of the mass function in open clusters.

In addition to best-fit values and uncertainties, we provide the full posterior chains from our Bayesian inference. These allow for reproducibility and enable users to reweight the posteriors using alternative priors (e.g., from independent metallicity constraints) for refined analyses.

This paper is structured as follows: we describe the MiMO method and fitting procedure in Section~\ref{sec:method}, present the resulting catalog in Section~\ref{sec:results}, discuss the results in detail and compare them with previous studies in Section~\ref{sec:discussion},  and summarize our conclusions in Section~\ref{sec:summary}.

\begin{table*}
\begin{center}
    \caption{Fitting Sample Selection Criteria.}
    \footnotesize
    \addtolength{\tabcolsep}{1.5pt}
    \begin{tabular}{lccccccccccccccccccccc}
    \hline
    \hline

Cluster & ra & dec & $\mu_\alpha^\ast$ & $\mu_\delta$ & $\sigma_{\mu_\alpha^\ast}$ & $\sigma_{\mu_\delta}$ & $\varpi_\mathrm{cl}$ & $\sigma_{\varpi,\mathrm{cl}}$ & $r_{50}$ & $N_{r_{50}}$ & $N_\mu$ & $N_\varpi$ & $N_{\varpi_{near}}$ & $N_{\mu_{fs}}$ \\
    \midrule

ASCC\_10 & $51.87$ & $34.981$ & $-1.737$ & $-1.368$ & $0.159$ & $0.143$ & $1.459$ & $0.1$ & $0.558$ & $3$ & $6$ & $4$ & $99$ & $8$ \\
ASCC\_101 & $288.399$ & $36.369$ & $0.934$ & $1.288$ & $0.205$ & $0.258$ & $2.488$ & $0.057$ & $0.372$ & $3$ & $8$ & $10$ & $99$ & $8$ \\
ASCC\_105 & $295.548$ & $27.366$ & $1.464$ & $-1.635$ & $0.162$ & $0.145$ & $1.783$ & $0.065$ & $0.648$ & $3$ & $6$ & $6$ & $99$ & $8$ \\
ASCC\_108 & $298.306$ & $39.349$ & $-0.519$ & $-1.69$ & $0.099$ & $0.129$ & $0.838$ & $0.048$ & $0.537$ & $3$ & $5$ & $4$ & $6$ & $8$ \\
ASCC\_11 & $53.056$ & $44.856$ & $0.926$ & $-3.03$ & $0.163$ & $0.147$ & $1.141$ & $0.061$ & $0.312$ & $3$ & $6$ & $4$ & $99$ & $8$ \\
ASCC\_110 & $300.742$ & $33.528$ & $0.271$ & $-3.132$ & $0.064$ & $0.05$ & $0.497$ & $0.036$ & $0.203$ & $3$ & $6$ & $6$ & $10$ & $8$ \\
ASCC\_111 & $302.891$ & $37.515$ & $-1.15$ & $-1.524$ & $0.151$ & $0.154$ & $1.166$ & $0.059$ & $0.537$ & $3$ & $6$ & $6$ & $99$ & $8$ \\
ASCC\_113 & $317.933$ & $38.638$ & $0.8$ & $-3.679$ & $0.125$ & $0.163$ & $1.762$ & $0.041$ & $0.529$ & $3$ & $8$ & $10$ & $99$ & $8$ \\
ASCC\_115 & $329.28$ & $51.558$ & $-0.549$ & $-0.543$ & $0.068$ & $0.088$ & $1.311$ & $0.034$ & $0.25$ & $3$ & $6$ & $6$ & $99$ & $8$ \\
ASCC\_12 & $72.4$ & $41.744$ & $-0.634$ & $-2.794$ & $0.181$ & $0.122$ & $0.941$ & $0.069$ & $0.303$ & $3$ & $6$ & $2$ & $10$ & $8$ \\
ASCC\_123 & $340.299$ & $53.986$ & $12.093$ & $-1.407$ & $0.473$ & $0.437$ & $4.262$ & $0.172$ & $1.294$ & $3$ & $6$ & $6$ & $99$ & $8$ \\
ASCC\_127 & $347.205$ & $64.974$ & $7.474$ & $-1.745$ & $0.26$ & $0.263$ & $2.633$ & $0.081$ & $0.627$ & $3$ & $6$ & $6$ & $99$ & $8$ \\
ASCC\_128 & $349.949$ & $54.435$ & $1.236$ & $0.186$ & $0.139$ & $0.114$ & $1.509$ & $0.058$ & $0.513$ & $3$ & $6$ & $6$ & $99$ & $8$ \\
ASCC\_16 & $81.198$ & $1.655$ & $1.355$ & $-0.015$ & $0.265$ & $0.248$ & $2.838$ & $0.104$ & $0.376$ & $3$ & $6$ & $6$ & $99$ & $8$ \\
ASCC\_19 & $81.982$ & $-1.987$ & $1.152$ & $-1.234$ & $0.252$ & $0.219$ & $2.768$ & $0.089$ & $0.605$ & $3$ & $6$ & $2$ & $99$ & $8$ \\
    $\ldots$\\

    \hline
    \hline
    \end{tabular}
\label{table:oc_select}
\end{center}
\tablecomments{This table lists the parameters used to define the data selection for each cluster, corresponding to the criteria in Section~\ref{sec:data_select}. Columns (2)-(9) list the mean cluster properties: right ascension (ra), declination (dec), mean proper motions ($\mu_{\alpha}^*$, $\mu_{\delta}$), their corresponding dispersions ($\sigma_{\mu_{\alpha}^*}$, $\sigma_{\mu_{\delta}}$), mean parallax ($\varpi_{cl}$), and its dispersion ($\sigma_{\varpi,cl}$). Column (10) is the half-number radius ($r_{50}$). The columns also list the multiplicative factors applied to the cluster's characteristic radius ($N_{r_{50}}$), proper motion dispersion ($N_{\mu}$), and parallax dispersion ($N_{\varpi}$). The column $N_{\varpi_{\text{near}}}$ defines an upper parallax limit to remove foreground stars, and is typically disabled by setting it to a large number. The column $N_{\mu_{fs}}$ defines the lower proper motion limit used to select the field star sample. The table shown here is a sample; the full version is available online.
}

\end{table*}

\begin{table}[htbp]
{\centering
\caption{Description and Prior Ranges of Parameters in MiMO}
\label{tab:model-paras}
\begin{tabular*}{1\columnwidth}{l @{\extracolsep{\fill}} ll}
    \hline
    \hline
     & Range &Description \\
    \midrule
    \multicolumn{3}{l}{\textit{Isochrone parameters}} \\
    \cmidrule(l){1-3}
    \quad logAge & $[6.2, 10.1]$& $\log_{10}$ cluster age (year)\\
    \quad  $\mathrm{DM}$ & $[3, 15]^\ast$ & distance modulus (mag)  \\
    \quad  $A_V$ & $[0, 3]$ & dust extinction at the $V$ band (mag) \\
    \quad  $\mathrm{[Fe/H]}$ & $[-2.1, 0.5]$ & $\log_{10}$ iron-to-hydrogen ratio\\
     & &   relative to the Sun (dex) \\
    \midrule
    \multicolumn{3}{l}{\textit{Mass function parameter}} \\
    \cmidrule(l){1-3}
    \quad  $\alpha_\mathrm{MF}$ & $[-4, 2]$& power-law index of Salpeter's MF \\
    \midrule
    \multicolumn{3}{l}{\textit{Binary parameters}$^\dagger$} \\
    \cmidrule(l){1-3}
    \quad  $f_\mathrm{b}$& $[0,1]$ & fraction of binaries$^\ddagger$ \\
    \midrule
    \multicolumn{3}{l}{\textit{Field parameter}$\dagger$} \\
    \cmidrule(l){1-3}
    \quad  $f_\mathrm{fs}$ & $[0, 1]$& fraction of field stars in the sample \\
    \hline
    \hline
\end{tabular*}}
\tablecomments{$^\ast$ The distance modulus range corresponds to distance from 40 pc to 10 kpc. 
$^\dagger$ The binary and field parameters are marginalized as nuisance parameters during inference.
$^\ddagger$ We only consider $f_\mathrm{b}$ for binary mass ratio $1\ge q\geq0.2$, because binaries with lower $q$ are nearly indistinguishable from single stars (see \citet{Li2020d}). 
}
\end{table}

\section{Method and Sample} 
\label{sec:method}

We briefly summarize the MiMO framework below and refer readers to \citetalias{Li2022k} for full details, including systematic validation with mock samples. We then describe the specific setup used in this work, including the observational sample, free parameters, and prior choices.

\subsection{Mixture model} \label{sec:MM}

MiMO models the observed number density distribution of a star sample in the CMD as a mixture of cluster members, $\phi_\mathrm{cl}$, and field stars, $\phi_\mathrm{fs}$,
\begin{align}
\label{eqn:phitot}
    \phi_\mix(m,c \mid \Theta) = (1-f_\mathrm{fs}) \phi_\mathrm{cl}(m,c \mid \Theta) + f_\mathrm{fs}\phi_\mathrm{fs}(m,c),
\end{align}
where $(m,c)$ denote the apparent magnitude and color, and $\Theta$ is the set of model parameters, including the fraction of field-star contamination in the sample, $f_\mathrm{fs}$.

The cluster component $\phi_\mathrm{cl}$ is modeled as a mixture of single stars and unresolved binaries, determined by the isochrone (age, metallicity, distance, extinction), stellar mass function, binary fraction, binary mass-ratio distribution, and observational errors. Specifically, we adopt PARSEC isochrones \citep{Bressan2012b} with the Gaia EDR3 photometric system \citep{Riello2021}, and the YBC extinction model \citep{Chen2019c}.%
\footnote{The isochrones are queried in batch using a script written by Zhaozhou Li (as part of \citet{Li2020d}), \url{https://github.com/syrte/query_isochrone}.}
Both the stellar mass function and the binary mass-ratio distribution are assumed to follow power-law forms, characterized by slopes $\amf$ and $\gq$, respectively.
$\phi_\mathrm{cl}$ is evaluated individually for each star, incorporating its photometric uncertainties and normalization of the selection function.

The field population $\phi_\mathrm{fs}$ is modeled empirically from an auxiliary sample of neighboring field stars for each cluster, assuming they represent the same population as the field contaminants within the cluster region.

Given a sample of $N$ stars, $D = \{m_i, c_i\}_{i=1}^{N}$, the posterior distribution of model parameters follows Bayes' theorem,
\begin{equation}\label{eq:pdf}
    p (\Theta \mid D) \propto 
        \pi(\Theta)\prod\nolimits_{i=1}^{N}\ \phi_\mix(m_i, c_i \mid \Theta),
\end{equation}
where $\pi(\Theta)$ denotes the prior distribution.
We perform the parameter inference using nested sampling \citep{Skilling2004a, Skilling2006}, as implemented in the \texttt{dynesty} package \citep{Speagle2020},\footnote{\url{https://github.com/joshspeagle/dynesty}} which provides weighted posterior samples and the Bayesian evidence. The latter can be used to assess the significance of the cluster component relative to a pure field population, a topic we defer to a separate analysis.

\subsection{Sample Selection}\label{sec:data_select}

MiMO adopts very inclusive sample selection criteria by employing a probabilistic mixture model to account for field-star contamination, rather than relying on a strictly selected member sample. This strategy improves both the statistical precision and robustness of parameter estimation.

For each OC, we select the fitting sample from the Gaia DR3 source catalog \citep{Vallenari2022}, following the procedure outlined in \citetalias{Li2022k}. The fiducal selection criteria are
\begin{equation}
\begin{aligned}
    &G < 18 \mathrm{mag}, \\
    &r < N_{r_{50}} \cdot r_{50}, \\
    &\varpi > \varpi_\mathrm{cl} - N_{\varpi} \cdot \sigma_{\varpi, \mathrm{cl}}, \\
    &\varpi > \varpi_\mathrm{cl} + N_{m_{\text{near}}} \cdot 
    \sigma_{\varpi, \mathrm{cl}}, \\
    &\Delta \mu < N_{\mu} \cdot \sigma_{\mu, \mathrm{cl}}.
\end{aligned}
\end{equation}

Here, $G$, $r$, and $\varpi$ are the magnitude, angular separation from the cluster center, and parallax for an individual star, respectively. The term $\Delta \mu = \sqrt{(\Delta \mu_\alpha^\ast)^2 + (\Delta \mu_\delta)^2}$ represents the deviation from the cluster's mean proper motion. The astrometric properties ($r_{50}$, $\varpi_\mathrm{cl}$, $\sigma_{\varpi,\mathrm{cl}}$, $\mu_\mathrm{cl}$, and $\sigma_{\mu, \mathrm{cl}}$) are adopted from \citet{Cantat-Gaudin2020b}.

The multiplicative factors ($N_{r_{50}}$, $N_{\mu}$, $N_{\varpi}$, $N_{\varpi_{\text{near}}}$) allow us to tune the selection volume for each cluster. Our fiducial, loose criteria correspond to ($N_{r_{50}}$, $N_{\mu}$, $N_{\varpi}$) = (3, 6, 6), with $N_{\varpi_{\text{near}}}$ typically set to a large value (e.g., 99) to effectively disable the foreground cut. The selection factors were adjusted on a case-by-case basis to optimize the input sample. For clusters in highly contaminated fields, we adopted more stringent criteria (i.e., smaller N factors) to reduce the number of field stars. Conversely, for certain clusters where the cataloged astrometric dispersions appeared underestimated, we relaxed the criteria (i.e., larger N factors) to ensure a higher completeness of member stars. The specific factors used for each OC are listed in Table \ref{table:oc_select}.

\citet{Li2020d} reported that the intrinsic dispersion of OC main sequences is broader than expected from Gaia's formal photometric uncertainties. To account for this, we add an additional 0.01 mag to the magnitude uncertainties for all stars in the fitting sample (also adopted by \citealt{Li2022k, liuPhotometricDeterminationUnresolved2025}).

As mentioned above, we construct a nonparametric empirical model to describe the distribution of field-star contamination in the CMD. For each cluster, we select stars from the same sky region, magnitude range, and parallax range as the fitting sample, but with proper motions that deviate significantly from the cluster mean: $\Delta \mu > 8\sigma_{\mu, \mathrm{cl}}$.

\subsection{Free Parameters}
The full set of free parameters used in MiMO are listed in Table~\ref{tab:model-paras}. These include isochrone parameters (age, metallicity, distance modulus, and extinction), the stellar mass function slope, and parameters for the binary and field star populations.

However, the determination of binary-related parameters is challenging due to known discrepancies between theoretical isochrone models and the observed main sequences of OCs \citep{Li2020d,wangEmpiricalColorCorrection2025,liuPhotometricDeterminationUnresolved2025}. As shown in \citet{Li2020d}, while these discrepancies do not significantly affect the inference of key isochrone parameters like age and distance, the derived binary properties are highly sensitive to the precise location of the isochrone.

Given this sensitivity, the current version of MiMO cannot provide robust constraints on the binary population. We therefore treat the binary fraction ($f_b$) as a nuisance parameter rather than a physically meaningful result in this study. For the same reason, we adopt a simplified model for the binary mass ratio distribution, assuming a fixed power-law index of $\gamma_{q}=0$. We acknowledge this is a simplification, as detailed studies have shown that binary mass ratio distributions are more complex, depending on the primary star's mass and orbital period (e.g., \citet{moeMindYourPs2017,liuMassdependentRadialDistribution2025a}).

More reliable estimates for binary parameters may be obtained by empirically correcting the isochrone to better match the observed main-sequence ridge line, for example using robust Gaussian processes \citep{liRobustGaussianProcess2021},\footnote{\url{https://github.com/syrte/robustgp}} an improvement we defer to future work.

\subsection{Choice of Priors} \label{sec:feh_prior}

In Bayesian inference, incorporating well-motivated priors helps reduce parameter degeneracies and improve the robustness of model fitting.

Since spectroscopic [Fe/H] measurements are more reliable than photometric estimates, we use literature spectroscopy values as priors where available \citep{Netopil2016a, Carrera2019a,donorOpenClusterChemical2020, 2021MNRAS.503.3279S, Fu2022c}. Specifically, we adopt a truncated Gaussian prior $\mathcal{N}(\mathrm{[Fe/H]_P}, \sigma_{\mathrm{[Fe/H]_P}})$ within the range $[-2.1, 0.5]$. When multiple measurements are available for a given cluster, we use their weighted mean as $\mathrm{[Fe/H]_P}$ and the corresponding standard deviation as $\sigma_{\mathrm{[Fe/H]_P}}$, following \citet{Schmidt2021}. For OCs without spectroscopic metallicity measurements, we adopt a Global Prior. This is a truncated Gaussian distribution, $\mathcal{N}(\mu=-0.063, \sigma=0.146)$, bounded within the range $[-2.1, 0.5]$. This prior is derived from the global distribution of spectroscopic metallicities of clusters in our sample, as detailed in Section~\ref{sec:catalog_compare}. This choice is more physically motivated than a simple uniform prior, centering the probability on typical OC metallicities while still allowing for the possibility of more extreme values.

Uniform priors are used for all other parameters, with the allowed ranges summarized in Table~\ref{tab:model-paras}.
In this work, we also provide the full posterior likelihood chains for all parameters, enabling future users to reweight the results with alternative priors of their choice.

\section{Catalog} \label{sec:results}

Our analysis is based on the input sample of 1743 clusters from D21, which provides a high-confidence list of existing OCs. Using this as our target list, we apply the MiMO with Gaia DR3 data to produce a homogeneous catalog. The primary products for each cluster include not only the fundamental physical parameters (age, metallicity, distance, and extinction), but also two key outputs derived self-consistently from our Bayesian framework: the stellar mass function slope and photometric membership probabilities for individual stars. After a thorough process of visual inspection and quality assessment of the fitting results, we produced a final, reliable catalog of 1232 clusters. 

As a practical note on the implementation, a full Bayesian analysis is computationally intensive. For a typical cluster fitting sample containing a few hundred stars, a complete MiMO run takes approximately 3 hours. This runtime increases to about 10 hours for larger samples consisting of several thousand stars. These benchmarks were performed on a standard modern CPU, and performance may vary depending on the hardware.

\subsection{Catalog of OC Parameters} \label{sec:oc_catalog}

 \begin{figure*}[!htbp]\label{fig:data_qlt}
  \centering
  \includegraphics[width=1\textwidth]{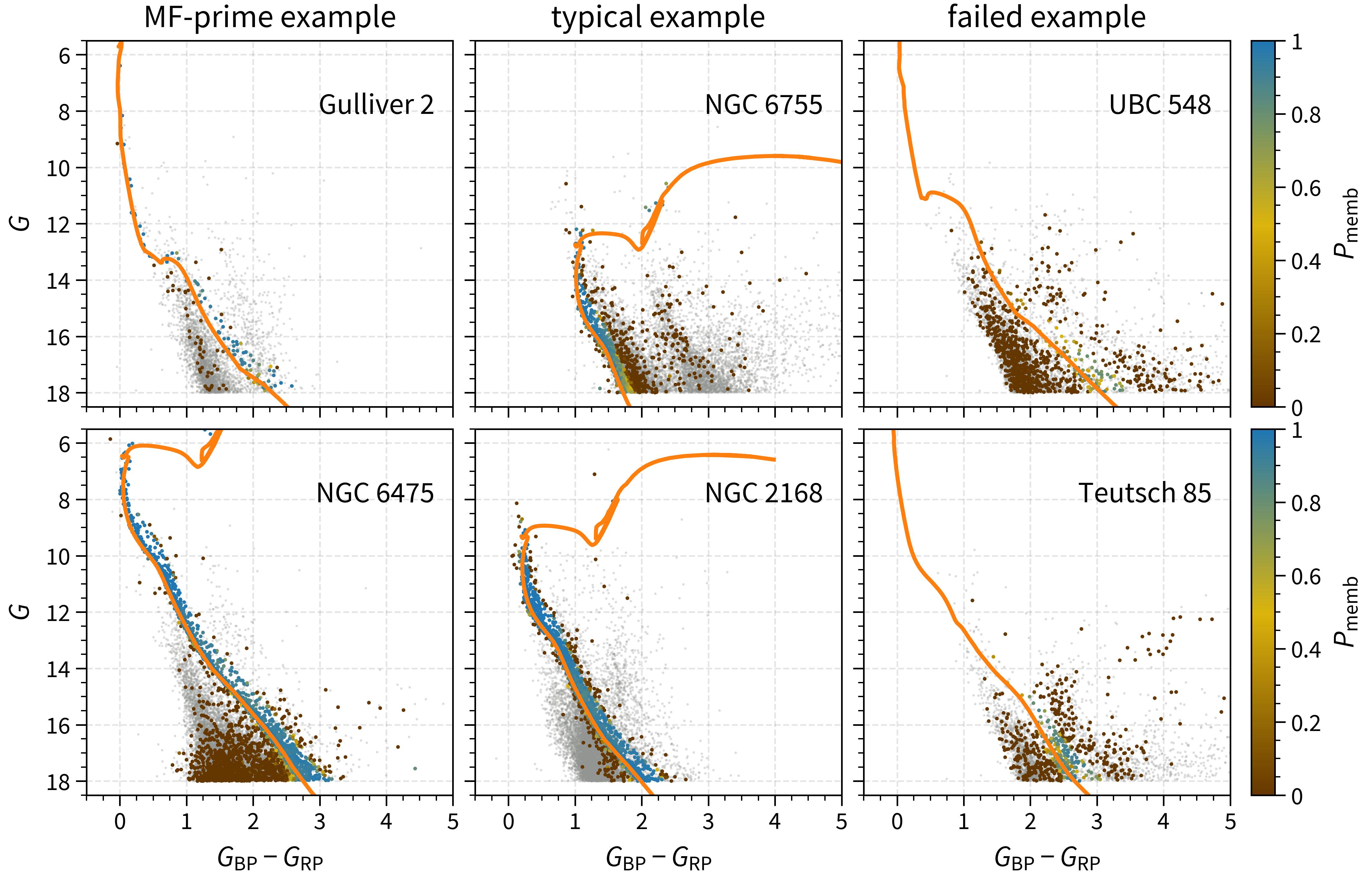}
  \caption{Illustration of MiMO on OCs with different data quality. The color bar shows the photometric membership probability for individual stars. The left panels show high-quality clusters (named as ``MF-prime'') with clear and thin main sequences, while the middle panels represent typical cases with moderate main-sequence broadening. The right panels display clusters where MiMO failed to fit a reasonable isochrone due to extreme field star contamination. In each panel, grey dots represent the accessory star sample used to construct the nonparametric field-star model, and the orange line indicates the best-fit isochrone from MiMO.}
\end{figure*}

Our final catalog provides a homogeneous set of parameters for all 1232 clusters. While we report the MF slope for every cluster, the accuracy of this parameter is highly sensitive to the quality of the CMD, particularly the width of the main sequence which can be broadened by effects like differential reddening.

Therefore, we identify an \textit{MF Prime sample} of 163 clusters for which the derived MF slope is considered most reliable. This selection is based on a visual inspection of the CMDs, prioritizing clusters that exhibit a narrow, well-defined main sequence. In our public data release, this subsample is marked with a "MF\_flag=1" for user guidance.

The final catalog provides a homogeneous determination of physical parameters for the 1232 OCs (see Table\ref{table:oc_cat}). Figure~\ref{fig:data_qlt} illustrates the visual basis for our quality flag. Clusters assigned to the "MF Prime sample" (left panels) are characterized by a narrow and well-defined main sequence. The middle panels show a typical case from the wider catalog. While the isochrone fit for fundamental parameters is still robust for these clusters, the inferred MF slope is less reliable. Significant main-sequence broadening(caused by effects not explicitly parameterized in our model, such as differential reddening or underestimated photometric observational error) can cause the model to accommodate these scattered stars by favoring a larger field contamination fraction. Consequently, the determination of MF slope will be biased. The right panels show clusters where MiMO failed to fit a plausible isochrone, typically due to extreme field-star contamination, which were manually excluded from the catalog.

\movetabledown=7cm
\begin{rotatetable*}
\begin{deluxetable*}{lcccccccccccccccccccccccl}
\setlength{\tabcolsep}{2.3pt}
\footnotesize
\tablecaption{Catalog of Inferred Parameters\label{tab:full_params}}
\tablehead{
{Cluster} & \multicolumn{4}{c}{$\log\text{Age (yr)}$} & \multicolumn{4}{c}{$[\text{Fe/H}]$ (dex)} & \multicolumn{4}{c}{DM (mag)} & \multicolumn{4}{c}{$A_V$ (mag)} & \multicolumn{4}{c}{$\alpha_{\text{MF}}$} & \multicolumn{2}{c}{$[\text{Fe/H}]$ prior} & {MF flag} \\
\cmidrule(lr){2-5} \cmidrule(lr){6-9} \cmidrule(lr){10-13} \cmidrule(lr){14-17} \cmidrule(lr){18-21} \cmidrule(lr){22-23}
 & best-fit & 50th & 16th & 84th & best-fit & 50th & 16th & 84th & best-fit & 50th & 16th & 84th & best-fit & 50th & 16th & 84th & best-fit & 50th & 16th & 84th & $\mu$ & $\sigma$ &
}
\startdata
ASCC\_10 & $8.586$ & $8.587$ & $8.572$ & $8.615$ & $-0.041$ & $-0.038$ & $-0.044$ & $-0.031$ & $8.993$ & $8.987$ & $8.974$ & $9.001$ & $0.691$ & $0.686$ & $0.668$ & $0.704$ & $-2.263$ & $-2.293$ & $-2.471$ & $-2.109$ & $-0.024$ & $0.011$ & $0$ \\
ASCC\_101 & $8.616$ & $8.626$ & $8.613$ & $8.640$ & $0.154$ & $0.151$ & $0.143$ & $0.160$ & $8.036$ & $8.036$ & $8.024$ & $8.047$ & $0.060$ & $0.049$ & $0.034$ & $0.064$ & $-1.661$ & $-1.868$ & $-2.045$ & $-1.697$ & $-0.063$ & $0.146$ & $1$ \\
ASCC\_105 & $8.159$ & $8.147$ & $8.104$ & $8.201$ & $0.053$ & $0.057$ & $0.051$ & $0.063$ & $8.651$ & $8.652$ & $8.641$ & $8.664$ & $0.373$ & $0.368$ & $0.353$ & $0.380$ & $-2.251$ & $-2.271$ & $-2.385$ & $-2.159$ & $0.045$ & $0.019$ & $0$ \\
ASCC\_108 & $8.360$ & $8.370$ & $8.319$ & $8.392$ & $-0.145$ & $-0.048$ & $-0.147$ & $0.052$ & $10.109$ & $10.183$ & $10.106$ & $10.262$ & $0.447$ & $0.426$ & $0.398$ & $0.451$ & $-2.948$ & $-2.904$ & $-2.988$ & $-2.793$ & $-0.106$ & $0.060$ & $0$ \\
ASCC\_11 & $8.622$ & $8.630$ & $8.608$ & $8.716$ & $-0.150$ & $-0.151$ & $-0.158$ & $-0.147$ & $9.477$ & $9.473$ & $9.461$ & $9.486$ & $0.765$ & $0.753$ & $0.732$ & $0.765$ & $-2.866$ & $-2.843$ & $-2.974$ & $-2.708$ & $-0.162$ & $0.027$ & $1$ \\
ASCC\_110 & $8.811$ & $8.799$ & $8.770$ & $8.821$ & $0.172$ & $0.161$ & $0.067$ & $0.233$ & $11.204$ & $11.198$ & $11.165$ & $11.271$ & $1.012$ & $1.021$ & $0.995$ & $1.131$ & $-2.214$ & $-2.174$ & $-2.544$ & $-1.827$ & $-0.063$ & $0.146$ & $0$ \\
ASCC\_111 & $8.424$ & $8.369$ & $8.325$ & $8.419$ & $0.149$ & $0.165$ & $0.146$ & $0.342$ & $9.721$ & $9.745$ & $9.717$ & $9.788$ & $0.609$ & $0.585$ & $0.463$ & $0.657$ & $-2.874$ & $-2.885$ & $-3.005$ & $-2.763$ & $-0.063$ & $0.146$ & $0$ \\
ASCC\_113 & $8.579$ & $8.581$ & $8.570$ & $8.625$ & $0.340$ & $0.251$ & $0.242$ & $0.342$ & $8.897$ & $8.842$ & $8.826$ & $8.895$ & $0.015$ & $0.021$ & $0.009$ & $0.049$ & $-2.184$ & $-2.190$ & $-2.302$ & $-2.103$ & $-0.063$ & $0.146$ & $1$ \\
ASCC\_115 & $8.350$ & $8.390$ & $8.350$ & $8.443$ & $0.077$ & $0.060$ & $-0.035$ & $0.092$ & $9.331$ & $9.316$ & $9.232$ & $9.356$ & $0.733$ & $0.737$ & $0.701$ & $0.778$ & $-2.369$ & $-2.486$ & $-2.746$ & $-2.241$ & $-0.063$ & $0.146$ & $0$ \\
ASCC\_12 & $8.579$ & $8.563$ & $8.473$ & $8.590$ & $-0.156$ & $-0.150$ & $-0.159$ & $-0.137$ & $10.003$ & $10.019$ & $9.998$ & $10.054$ & $0.878$ & $0.890$ & $0.868$ & $0.928$ & $-2.872$ & $-2.877$ & $-3.043$ & $-2.722$ & $-0.162$ & $0.047$ & $0$ \\
ASCC\_123 & $7.782$ & $7.781$ & $7.774$ & $7.790$ & $-0.047$ & $-0.048$ & $-0.056$ & $-0.040$ & $7.004$ & $7.030$ & $6.994$ & $7.073$ & $0.220$ & $0.246$ & $0.220$ & $0.276$ & $-1.742$ & $-1.808$ & $-1.912$ & $-1.712$ & $-0.063$ & $0.146$ & $0$ \\
ASCC\_127 & $7.327$ & $7.325$ & $7.320$ & $7.330$ & $0.246$ & $0.251$ & $0.239$ & $0.259$ & $7.910$ & $7.909$ & $7.900$ & $7.917$ & $0.539$ & $0.535$ & $0.521$ & $0.548$ & $-2.231$ & $-2.194$ & $-2.302$ & $-2.087$ & $-0.063$ & $0.146$ & $1$ \\
ASCC\_128 & $8.023$ & $7.997$ & $7.975$ & $8.030$ & $-0.052$ & $-0.050$ & $-0.067$ & $-0.037$ & $9.156$ & $9.147$ & $9.111$ & $9.183$ & $0.624$ & $0.628$ & $0.613$ & $0.644$ & $-2.051$ & $-2.185$ & $-2.317$ & $-2.036$ & $-0.063$ & $0.146$ & $1$ \\
ASCC\_16 & $7.126$ & $7.127$ & $7.125$ & $7.130$ & $-0.041$ & $-0.044$ & $-0.047$ & $-0.040$ & $7.618$ & $7.613$ & $7.606$ & $7.619$ & $0.141$ & $0.147$ & $0.136$ & $0.153$ & $-1.979$ & $-1.987$ & $-2.034$ & $-1.932$ & $0.004$ & $0.018$ & $0$ \\
ASCC\_19 & $7.126$ & $7.123$ & $7.121$ & $7.125$ & $-0.141$ & $-0.146$ & $-0.148$ & $-0.142$ & $7.375$ & $7.382$ & $7.378$ & $7.389$ & $0.152$ & $0.160$ & $0.151$ & $0.165$ & $-2.012$ & $-1.997$ & $-2.035$ & $-1.961$ & $-0.053$ & $0.034$ & $0$ \\
$\ldots$\\
\enddata
\tablecomments{
Column descriptions: (1) Cluster Name. (2-5) Logarithm of age (yr). (6-9) Metallicity. (10-13) Distance Modulus. (14-17) V-band extinction. (18-21) Slope of the stellar mass function. For each parameter group, we list the best-fit value, followed by the 50th (median), 16th, and 84th percentiles of the posterior distribution. (22-23) Mean ($\mu$) and standard deviation ($\sigma$) of the spectroscopic metallicity prior used. (24) MF Flag: 1 for clusters in the reliable "MF Prime" sample, 0 otherwise.
}
\label{table:oc_cat}
\end{deluxetable*}
\end{rotatetable*}

 \begin{figure}[!htbp]\label{fig:paras_distr}
  \centering
  \includegraphics[width=1\columnwidth]{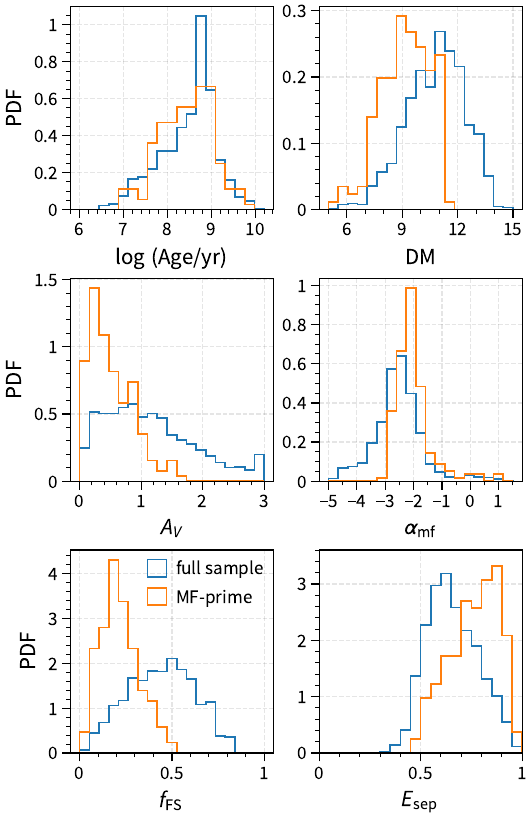}
  \caption{Parameter distributions of OCs in the catalog, for the full catalog and MF-prime samples, respectively.}
\end{figure}

 \begin{figure}[!htbp]\label{fig:mimo_uncer}
  \centering
  \includegraphics[width=1\columnwidth]{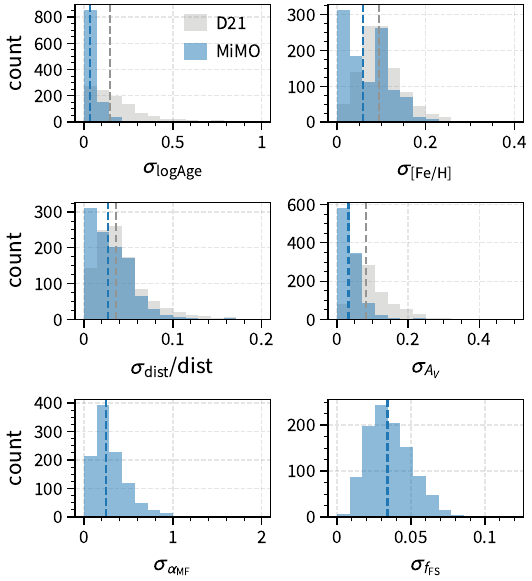}
  \caption{
  Distributions of formal uncertainties for parameters of OCs in our catalog, as inferred by MiMO.
  For reference, the uncertainty distributions from the \citetalias{Dias2021a} catalog are also shown when applicable.
}
\end{figure}

Figure~\ref{fig:paras_distr} compares the parameter distributions for the full catalog (blue histograms) against those for the MF Prime sample (orange histograms). The prime sample is biased towards clusters with smaller distances and extinctions, which contributes to their higher-quality CMDs and more reliable MF slope measurements. Both populations cover a similar age range, indicating that the quality of the MF determination is not strongly dependent on cluster age.

We further present the distribution of the formal errors (i.e., 68\% confidence regions returned by Bayesian inference) of our inferred parameters in Figure~\ref{fig:mimo_uncer}. For reference, the uncertainty distributions from the D21 catalog are also shown. The comparison demonstrates that the parameters derived using MiMO for our full catalog generally have smaller formal errors than those reported by D21. We emphasize that these formal errors from MiMO have been carefully validated through mock tests (\citetalias{Li2022k}) and thus represent reliable estimates of the true uncertainties.

\subsection{Photometric Membership Probability} \label{sec:memb_prob}

In addition to cluster parameters, MiMO provides the photometric membership probability, $p_{\text{memb}}$, for individual stars in the input sample for each OC (see Table~\ref{table:p_memb}). 
These probabilities are computed based on the best-fit isochrone model of the cluster.

The photometric membership probability offers an independent assessment of cluster membership, complementary to kinematic probabilities derived from astrometric data. Combining these independent probabilities enables a more robust member classification, which effectively reduces field contamination and enhances the reliability of subsequent analyses.

Another important application of $p_{\text{memb}}$ is the identification of anomalous objects such as blue stragglers. Stars with high kinematic probabilities but low photometric $p_{\text{memb}}$ may deviate from standard isochrone evolution, flagging them as potential blue straggler candidates. A dedicated follow-up study will present a catalog of blue straggler candidates based on these combined criteria.

We further demonstrate the utility of $p_{\text{memb}}$ in quantifying the degree of separation between the distributions of cluster and field populations in the CMD. We define the separation index $E_\mathrm{sep}$ of a cluster, originally introduced by \citet{Shao1996},
\begin{equation}
E_{\mathrm{sep}} = 1 - \frac{N  \sum_{i} [p_{\mathrm{memb},i}  (1 - p_{\mathrm{memb},i})]}{\left( \sum_{i} p_{\mathrm{memb},i} \right)  \left( \sum_{i} (1 - p_{\mathrm{memb},i}) \right)},
\end{equation}
where $p_{\mathrm{memb},i}$ denotes the photometric membership probability of the $i$-th star, and $N$ is the number of stars. By construction,  $E_\mathrm{sep} = 1$ corresponds to perfect separation between the cluster and field populations in the CMD ($p_{\mathrm{memb},i}$ being either 0 or 1), while $E_\mathrm{sep} = 0$ indicates complete overlap  ($p_{\mathrm{memb},i}=0.5$, thus indistinguishable).

\begin{figure}[!htbp]\label{fig:e_mix}
  \centering
  \includegraphics[width=0.85\columnwidth]{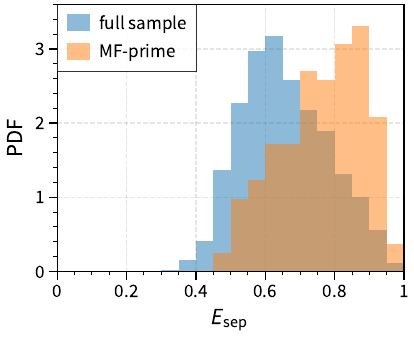}
  \caption{Distribution of the separation index, $E_\mathrm{sep}$, for the full catalog and MF-prime samples, respectively. $E_\mathrm{sep}$ indicates the level of separation between the distributions of cluster and field populations in the CMD.}
\end{figure}

Figure~\ref{fig:e_mix} shows the distribution of $E_\mathrm{sep}$ for our catalog. As expected, clusters in the MF Prime sample exhibit systematically higher $E_\mathrm{sep}$ values than the full catalog population. This confirms that the prime sample clusters, selected for their clean CMDs, also have a more distinct and separable population of member stars from the field.

\begin{table}
\begin{center}
    \caption{Photometric Membership Probabilities for Individual Stars in Each OC}
    \small\addtolength{\tabcolsep}{-2pt}
    \begin{tabular}{ccccc}
    \hline
    \hline

Cluster & Source ID & RA & Dec & $P_\mathrm{memb}$ \\
    \midrule

ASCC\_101 & 2050390247223041408 & 288.142 & 35.297 & 0.997 \\
ASCC\_101 & 2050401654656252160 & 288.142 & 35.532 & 0.989 \\
ASCC\_101 & 2050402101332868352 & 288.168 & 35.561 & 0.979 \\
ASCC\_101 & 2050446597192533376 & 289.297 & 35.547 & 0.992 \\
ASCC\_101 & 2050447112590439808 & 289.373 & 35.598 & 0.000 \\
ASCC\_101 & 2050464704774735232 & 289.193 & 35.783 & 0.000 \\
ASCC\_101 & 2050469927456762112 & 289.432 & 35.900 & 0.000 \\
ASCC\_101 & 2050486454489669888 & 288.929 & 35.666 & 0.937 \\
ASCC\_101 & 2050486488849410688 & 288.941 & 35.678 & 0.000 \\
ASCC\_101 & 2050492703657655552 & 288.432 & 35.540 & 0.000 \\
ASCC\_101 & 2050497140368635136 & 288.350 & 35.641 & 0.992 \\
    $\ldots$\\

    \hline
    \hline
    \end{tabular}
\label{table:p_memb}
\end{center}
\end{table}

\section{Discussion} \label{sec:discussion}

In this section, we first discuss the influence of the adopted metallicity prior on our parameter estimates. We then present a detailed comparison of our results with previous catalogs.

\subsection{Influence of prior for metalicity} \label{sec:discussion_prior}

In Bayesian inference, the choice of prior can be crucial. When available, spectroscopic metallicity measurements provide the most precise constraints and are thus ideal as priors. However, such measurements exist for only a subset of clusters. It is therefore important to assess whether reliable and consistent parameter estimates can be obtained using alternative priors for clusters lacking spectroscopic data.

To investigate this, we tested three different priors using the subset of clusters with spectroscopic metallicities as reference:

\begin{itemize}[leftmargin=\parindent]
    \item (a) \emph{Spectroscopic Prior:} A truncated Gaussian distribution with mean $\mathrm{[Fe/H]}_P$ and standard deviation $\sigma_{\mathrm{[Fe/H]}_P}$, bounded within $[-2.1, 0.5]$.
    \item (b) \emph{Global Prior:} A Gaussian distribution with mean $-0.063$ and standard deviation $0.146$, derived from the global distribution of spectroscopic metallicities across the subset, truncated to the same interval.
    \item (c) \emph{Uniform Prior:} A flat distribution over the full metallicity range $[-2.1, 0.5]$.
\end{itemize}

\begin{figure}[!htbp]\label{fig:prior_effect}
  \centering
  \includegraphics[width=1\columnwidth]{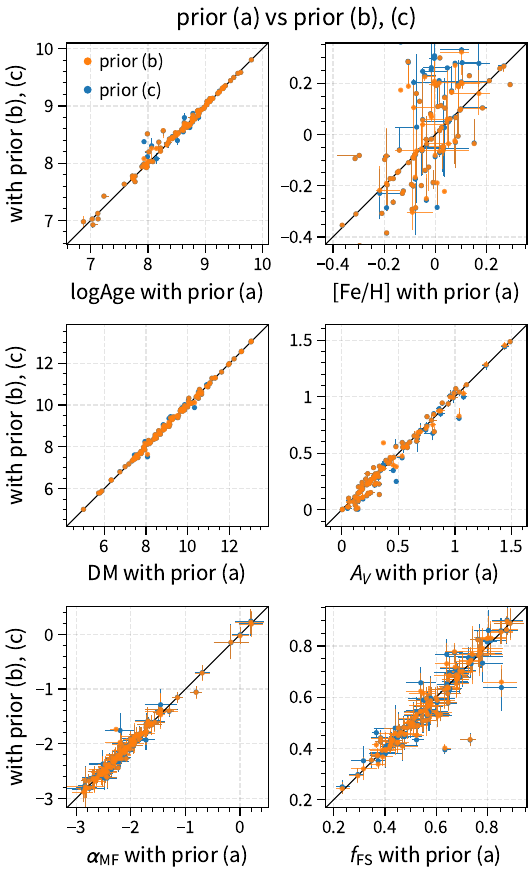}
  \caption{Comparison of cluster parameter estimates using different metallicity priors. Results with the Spectroscopic Prior (a) are shown on the horizontal axes. The orange points represent the Global Prior (b), and the blue points represent the Uniform Prior (c), both shown on the vertical axes.}

\end{figure}

Figure~\ref{fig:prior_effect} compares the resulting parameter estimates under these three priors. We find that, aside from [Fe/H] itself, the choice of metallicity prior has negligible impact on other inferred parameters, with all priors yielding consistent results. As expected, the global prior (b) and uniform prior (c) lead to greater scatter in individual [Fe/H] estimates compared to the spectroscopic prior, but they do not introduce systematic bias.

These results confirm that MiMO provides robust parameter estimates even in the absence of spectroscopic metallicity constraints. Accordingly, we adopt prior (a) for clusters with available spectroscopic metallicities, and the global prior (b) for the rest.

\

 \begin{figure*}[!htbp]\label{fig:compare_d21}
  \centering
  \includegraphics[width=1\linewidth]{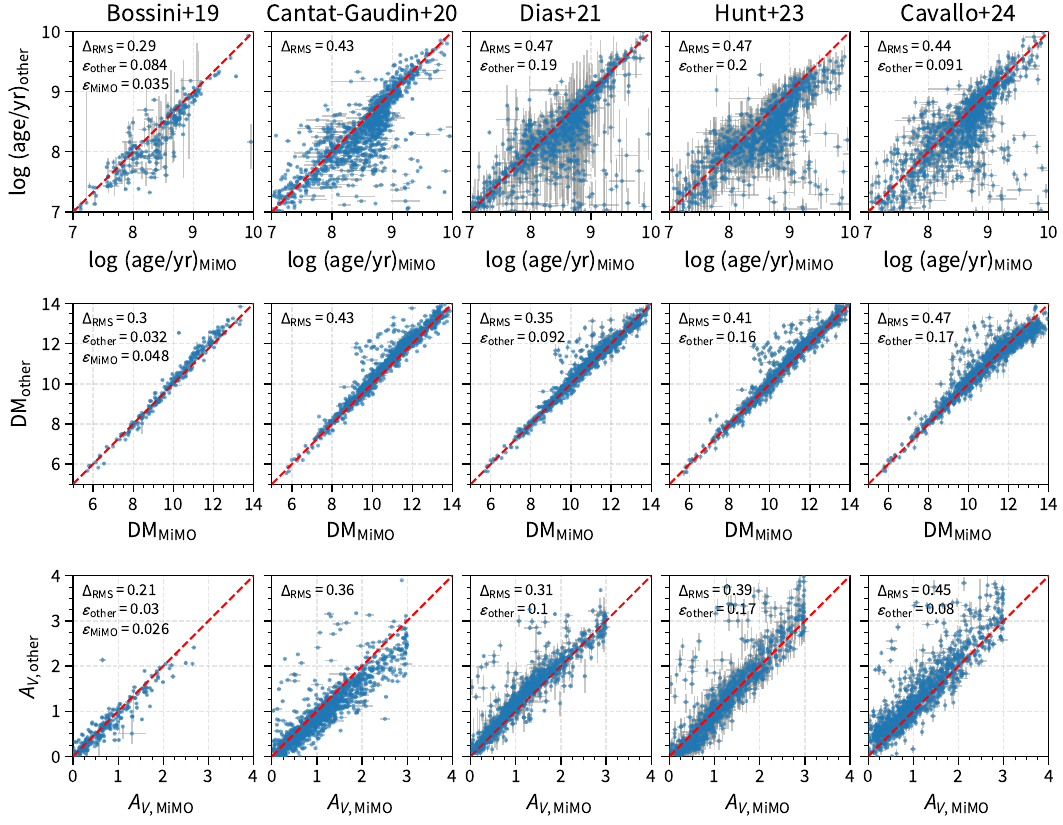}
  \caption{
  Comparison of age, distance modulus, and $A_V$ (rows) measured by MiMO ($x$-axis) versus \citet{Bossini2019a,Cantat-Gaudin2020b, Dias2021a, hunt2023} and \citet{cavalloParameterEstimationOpen2024}  (columns). Each point represents a star cluster. Error bars indicate the reported fitting uncertainties for each method (except for \citealt{Cantat-Gaudin2020b}, which does not provide error bars). The root-mean-square difference between the best fits, $\Delta_\mathrm{RMS}$, and the mean fitting uncertainties of each catalog, $\epsilon_\mathrm{MiMO}$ and $\epsilon_\mathrm{other}$, are indicated in the upper-right corner of each panel.
  }
\end{figure*}

\subsection{Comparison with previous results} \label{sec:catalog_compare}

We compare the parameters inferred by MiMO with several widely used catalogs in the literature 
\citep{Bossini2019a,Cantat-Gaudin2020b, Dias2021a, hunt2023,cavalloParameterEstimationOpen2024}. Figure~\ref{fig:compare_d21} presents a comparison of key cluster parameters, including age, distance, and extinction. 
Overall, there is good agreement between the different catalogs, particularly in cluster distances and extinctions. However, the uncertainties associated with MiMO estimates are generally smaller than those reported in the literature (see Figure~\ref{fig:mimo_uncer} for a comparison with \citealt{Dias2021a}).

Clusters showing larger discrepancies typically have lower data quality, either due to a small number of members or because the main sequence lies close to the field-star population. 
We note an apparently better agreement with the Bayesian method of \citet{Bossini2019a}, but this is largely because their analysis focuses only on clusters with high-quality data, which naturally leads to better consistency.

Notable discrepancies in age estimates are observed for some clusters. As discussed in \citetalias{Li2022k}, these differences likely arise from contrasting sample selection strategies (purity vs. completeness): \citetalias{Dias2021a} restricts its input to stars with high membership probabilities, whereas MiMO adopts a more inclusive selection criterion, resulting in higher completeness of member stars. 
This higher completeness improves both the precision and robustness of the derived parameters. For example, accidental exclusion of turnoff stars through overly stringent sample selection can lead to significant biases.

\begin{figure}[!htbp]\label{fig:comp_cmd}
  \centering
  \includegraphics[width=1\columnwidth]{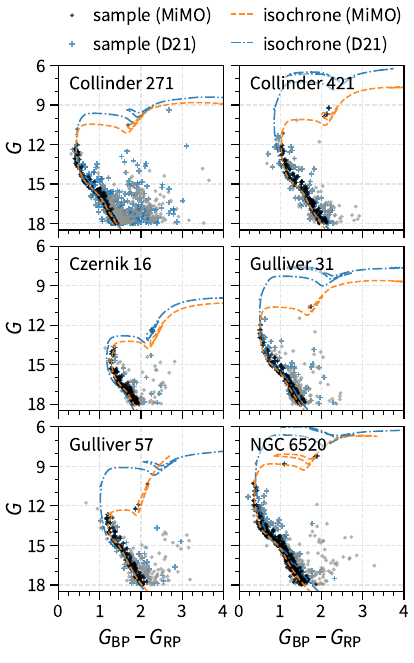}
  \caption{CMDs for six example clusters that show significant differences in age estimates between MiMO and D21. Small dots represent the MiMO input sample, with likely field stars ($p_\mathrm{memb} < 1 - f_\mathrm{fs}$) shown in gray, while crosses indicate the sample used by D21.
}
\end{figure}

Figure~\ref{fig:comp_cmd} further illustrates these differences by showing CMDs for six example clusters exhibiting significant discrepancies in age estimates between MiMO and \citetalias{Dias2021a}. Visual inspection suggests that the isochrones fitted by MiMO generally trace the data more closely, particularly near the turnoff region, supporting the reliability of MiMO.

A more thorough investigation of these discrepancies and the associated systematics is left to future work.

\section{Conclusion}
\label{sec:summary}
The Mixture Model for Open Clusters (MiMO; \citetalias{Li2022k}) is a robust and versatile Bayesian framework for accurately determining the physical parameters of OCs. Unlike traditional methods that rely on pre-selected high-probability members, MiMO explicitly models field star contamination, enabling the inclusion of a more complete set of stars in the fitting process. This approach significantly enhances the precision and robustness of the inferred parameters, which can be particularly critical for cluster age determinations.

In this work, we applied MiMO to Gaia DR3 photometric data, producing a comprehensive catalog for 1232 open clusters. We provide a homogeneous determination of their fundamental physical parameters and, crucially, the stellar mass function (MF) slope for every cluster in the catalog. Furthermore, we identify an ''MF-prime sample'' of 163 clusters. The high data quality of these clusters yields the most reliable MF measurements. This prime sample offers a solid foundation for future investigations into the evolution of the mass function in open clusters. Our results are consistent with existing OC catalogs but exhibit overall improved precision.

In a future work, we plan to extend MiMO by incorporating extinction dispersion (i.e., differential reddening) as a free parameter, enabling statistical studies of dust distribution within clusters and further improving the accuracy of derived physical properties. In addition, an empirical correction of the isochrone to match the observed main-sequence ridge line \citep{Li2020d,liuPhotometricDeterminationUnresolved2025} will improve constraints on binary star properties, which are highly sensitive to the exact location of an isochrone.

The full catalog described in this work is publicly available on the National Astronomical Data Center China-VO Paper-Data service: DOI: 10.12149/101693. The source code of MiMO is also available online at GitHub \footnote{\url{https://github.com/luly42/mimo}} with a copy, which includes the model isochrone files, deposited with the full catalog on China-VO. We hope it may serve as a valuable resource for the community and encourage users to adapt MiMO to their specific research needs.

\section*{Acknowledgments}
We thanks Dr.~He Zhao and Prof.~Chao Liu for helpful discussions. This work is supported by the National Natural Science Foundation of China (NSFC) under grant No. 12303026, 12273091, 12203100 and U2031139; the Science and Technology Commission of Shanghai Municipality (Grant No. 22dz1202400); the science research grants from the China Manned Space Project with No. CMS-CSST-2021-A08. This work was also sponsored by the Young Data Scientist Project of the National Astronomical Data Center and the Program of Shanghai Academic/Technology Research Leader. 
ZZL acknowledges the Marie Skłodowska-Curie Actions Fellowship under the Horizon Europe programme (101109759, ``CuspCore'').

This work has made use of data from the European Space Agency (ESA) mission
{\it Gaia} (\url{https://www.cosmos.esa.int/gaia}), processed by the {\it Gaia}
Data Processing and Analysis Consortium (DPAC,
\url{https://www.cosmos.esa.int/web/gaia/dpac/consortium}). Funding for the DPAC
has been provided by national institutions, in particular the institutions
participating in the {\it Gaia} Multilateral Agreement.

Data resources are supported by China National Astronomical Data Center (NADC) and Chinese Virtual Observatory (China-VO). This work is supported by Astronomical Big Data Joint Research Center, co-founded by National Astronomical Observatories, Chinese Academy of Sciences and Alibaba Cloud.

\bibliographystyle{aasjournal}
\bibliography{main}

\end{CJK*}
\end{document}